\documentclass[12pt]{article}
\usepackage[margin=1in]{geometry}
\usepackage{booktabs}
\usepackage{array}
\usepackage{amsmath}
\usepackage{graphicx}
\usepackage{threeparttable}
\usepackage{setspace}
\usepackage{xcolor}
\usepackage{float}
\title{Does Entry of Food-and-Drink Establishments Raise Local House Prices? Event-Study Evidence from London\thanks{%
* Corresponding author: Rong Zhao (\texttt{rong.zhao.25@ucl.ac.uk}).}}
\author{Wanqi Liu\textsuperscript{a} \and Rong Zhao\textsuperscript{a,*}}
\date{
\small
\textsuperscript{a} Centre for Advanced Spatial Analysis, University College London, London, UK
}

\begin{document}

\maketitle

\begin{abstract}
Restaurants, cafes, pubs, and takeaways are among the most visible markers of neighborhood change, yet whether their arrival is capitalised into nearby housing values remains empirically unsettled. We assemble a London-wide panel linking Land Registry prices, non-domestic EPC lodgement timings for food-and-drink establishments, and neighborhood amenity measures at the LSOA level. Our preferred annual event-study design defines treatment as the first clean-onset year in which an LSOA records at least two eligible EPC lodgements for food-and-drink establishments, after a two-year lookback with no prior entries. In this specification, pre-trend tests are not rejected in either the stacked or Sun-Abraham estimators, and log house prices rise gradually from about 0.5\% in the event year to roughly 3.4--3.7\% by years four and five. The results are consistent with local amenity capitalization following commercial entry, while remaining appropriately cautious about endogenous siting and concurrent redevelopment.
\end{abstract}


\section{Introduction}

Entry of restaurants, cafes, pubs, and takeaways is one of the most visible margins of neighborhood change. These establishments can raise local consumption value and alter how an area is perceived, but they also tend to open where demand is already strengthening. That makes commercial entry a natural object of interest in urban economics and a difficult object of causal measurement. If nearby house prices capitalize such entry, then commercial zoning, licensing, and conversion policy have distributional consequences that extend well beyond the business sector itself.

This paper studies that question in Greater London using an annual LSOA-year panel and the staggered timing of entry of food-and-drink establishments identified from non-domestic Energy Performance Certificate (EPC) lodgements. The core empirical challenge is endogenous siting: the same local fundamentals that attract new venues may already be pushing house prices up. We therefore use a deliberately narrow treatment definition. In the preferred design, an LSOA is first treated when it records at least two eligible lodgements for food-and-drink establishments after a two-year lookback window with no prior eligible entry. In this clean-onset event study, pre-trend tests are not rejected in either the stacked or Sun--Abraham estimators, and local log house prices rise gradually to roughly 3.4--3.7\% by years four and five.

The paper makes three contributions. First, it brings opening-timing data into the urban amenity and house-price literature by using EPC lodgements to track discrete local commercial entry rather than relying only on slowly moving amenity stocks. Second, it shows that treatment construction matters substantively: a clean-onset rule yields a more defensible dynamic design than broader entry definitions that mix new shocks with ongoing local churn. Third, it provides evidence from London, where dense mixed-use neighborhoods make local commercial change especially likely to be priced into nearby housing demand.

Our analysis speaks to two related literatures. A first literature studies gentrification and neighborhood upgrading through shifts in consumption spaces, retail landscapes, and symbolic markers of class change \cite{glass1964, smith1979, zukin1982, butler2003, hackworth2002, zukin2009}. This work makes clear that restaurants, cafes, and related venues are central to how urban change is experienced on the ground, but it is often less precise about whether those visible commercial shifts independently move nearby housing values.

A second literature estimates how local shocks are capitalized into property prices, including neighborhood revitalization, transport improvements, and other place-based interventions \cite{guerrieri2013, heckert2012, dube2014, dong2021, liu2023}. What remains relatively scarce is evidence on whether ordinary commercial entry itself, observed with usable timing at small spatial scale, is followed by local house-price appreciation. That question is empirically demanding because commercial entry is both neighborhood-facing and endogenous. We therefore keep broader spatial and stock-based commercial evidence in a supporting role and center the paper on a single clean-onset event-study design around local entry timing.

The remainder of the paper proceeds as follows. Section 2 describes the London setting, the EPC-based treatment measure, and the construction of the annual LSOA panel. Section 3 presents the empirical design and the main event-study evidence. Section 4 discusses interpretation and limitations. Section 5 concludes.


\section{Setting, Data, and Treatment}

\subsection{Urban Setting and Commercial Margin}

This study focuses on Greater London, where local high streets, mixed-use corridors, and neighborhood retail clusters are tightly linked to residential demand and redevelopment pressure. Food-and-drink establishments are an informative commercial margin because they are neighborhood-facing, frequent, and visible. Compared with larger-format retail or office uses, they enter in smaller local waves and plausibly affect day-to-day consumption opportunities, street activity, and neighborhood reputation at short spatial range. If local housing markets capitalize nearby amenity change, these are among the establishments most likely to matter.

At the same time, entry of food-and-drink establishments is not a randomly assigned shock. Businesses open where operators anticipate sufficient demand, where landlords can assemble suitable space, and where local commercial change is already underway. That makes this a substantively important but empirically difficult margin. The design is built around that tension: it uses the timing of discrete local entry waves, rather than cross-sectional amenity stocks, to study whether nearby house prices subsequently revalue.

\subsection{Core Data and Outcome}

We use Lower Layer Super Output Areas (LSOAs) as the geographic unit---small, stable areas of approximately 1,500 residents \cite{barton2016}. The main empirical design uses an annual LSOA-year panel covering 2008--2023. Our outcome is annual log house price, constructed by aggregating Land Registry transaction-price information to the LSOA-year level.

House price data come from the Land Registry's Price Paid Dataset. The treatment data come from the UK government's non-domestic EPC Register, which records certificate lodgements for commercial premises. Under the EPC regime in England and Wales, a non-domestic certificate is generally required when a property is constructed, sold, or let. For small business premises, lodgements therefore tend to cluster around occupancy changes and market entry, giving materially sharper timing information than stock-based amenity sources even though they do not isolate pure openings mechanically. We also merge supporting neighborhood covariates and confound checks, including deprivation, population density, transport accessibility, green-space access, and planning starts. Table \ref{tab:data_sources} summarizes the core data inputs used in the annual panel.

For the event study, we classify EPC records as belonging to food-and-drink establishments when the property-type field indicates restaurant, cafe, drinking establishment, pub/public house, or takeaway activity. The measure should therefore be read as a timing proxy for local commercial entry or reactivation. It remains imperfect---a lodgement can also reflect refurbishment, reletting, or sale-related activity---which is why the paper places weight on treatment construction and pre-trend behavior rather than on raw lodgement counts alone.

\begin{table}[H]
\centering
\caption{Data Sources and Coverage}
\label{tab:data_sources}
\begin{threeparttable}
\small
\begin{tabular}{@{}llll@{}}
\toprule
\textbf{Dimension} & \textbf{Source} & \textbf{Temporal Coverage} & \textbf{N (LSOAs)} \\
\midrule
House Prices & Land Registry PPD & Annual (2008--23) & 4,831 \\
Commercial Entry & EPC Register (non-domestic) & Annual (2008--23) & 4,831 \\
Planning Starts & Planning London Datahub & Annual (2008--23) & 4,831 \\
Deprivation and Demography & IMD + ONS/Census sources & Mostly static & 4,831 \\
Transport and Green Access & TfL + OSM-derived measures & Mostly static & 4,831 \\
Commercial Context Measures & OSM + Google Places API & Supporting only & 4,831 \\
\bottomrule
\end{tabular}
\begin{tablenotes}
\item Notes: All datasets are harmonized to 2011 LSOA boundaries. The headline results use the annual LSOA-year panel built from house prices and EPC-based treatment timings; other measures are used as supporting covariates or confound checks.
\end{tablenotes}
\end{threeparttable}
\end{table}

\subsection{Treatment Construction}

Treatment is assigned at the LSOA-year level. Our preferred specification labels an LSOA as treated in the first year it records at least two eligible EPC lodgements for food-and-drink establishments, provided the preceding two years contain no prior eligible establishment entry. This threshold-and-lookback rule is designed to capture a discrete local entry wave rather than ongoing commercial churn and to reduce contamination from repeat lodgements attached to the same local commercial stock. Under this definition, the preferred design contains 774 treated-role LSOAs.

\begin{figure}[H]
\centering
\includegraphics[width=\textwidth]{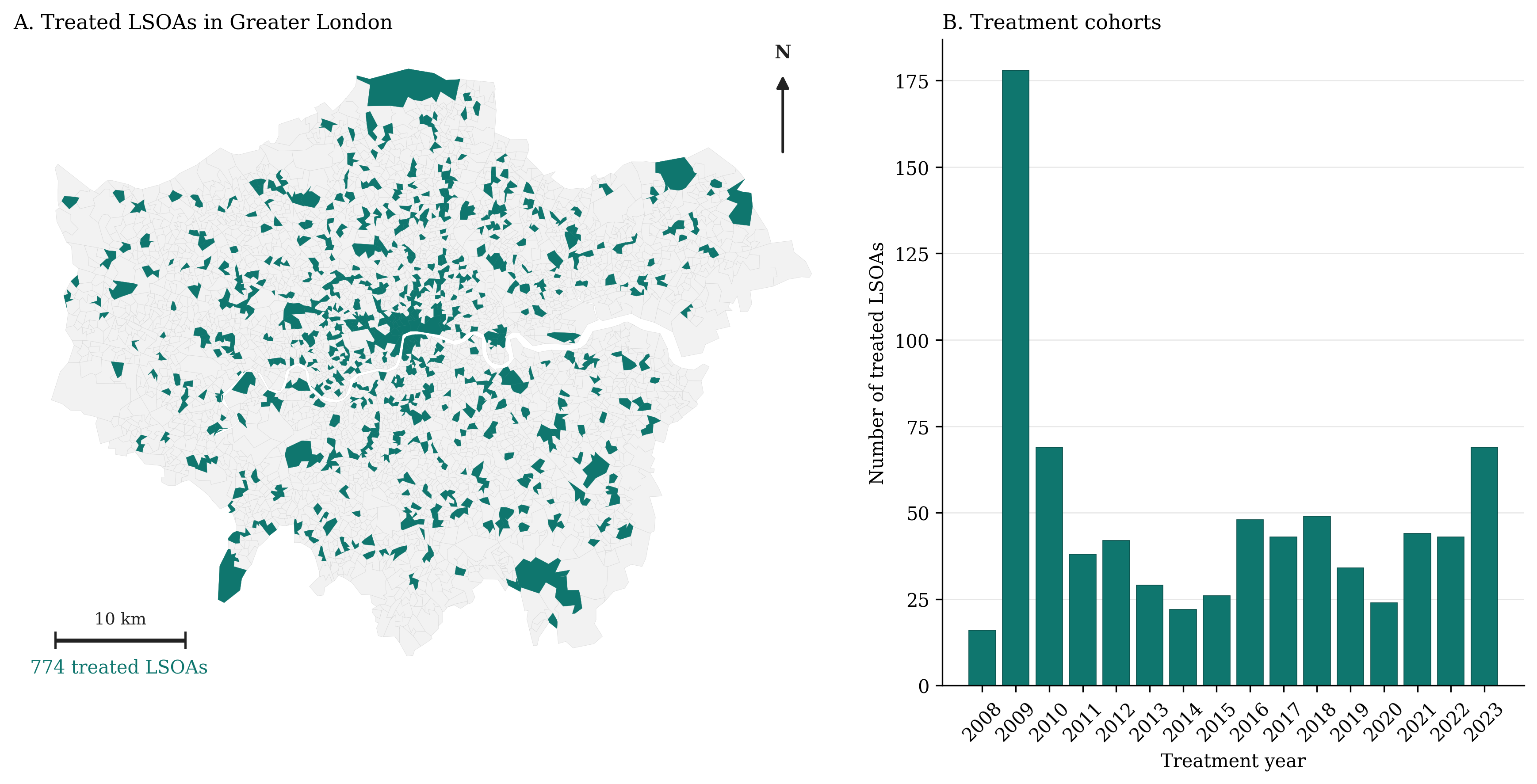}
\caption{Geographic and cohort structure of the preferred clean-onset treatment definition. Panel A maps the 774 treated LSOAs in the preferred clean-onset design for entry of food-and-drink establishments and includes a north arrow and 10 km scale bar for orientation. Panel B reports the number of treated LSOAs by treatment year, showing support across sixteen cohorts from 2008 to 2023.}
\label{fig:setting}
\end{figure}

Figure \ref{fig:setting} shows the spatial and temporal footprint of this preferred design. Treated LSOAs are distributed across Greater London rather than being confined to a single cluster, though they are denser in the inner-city areas where commercial turnover is most active. Treatment timing also spans sixteen cohorts from 2008 to 2023. The 2009 cohort is the largest, but the right tail remains substantively populated, which matters for the medium-run event-study coefficients reported below.

\subsection{Supporting Neighborhood Measures}

We retain broader commercial-context and neighborhood measures in a supporting role. In particular, retail and amenity-stock indicators derived from OpenStreetMap and Google Places are useful for describing London's commercial geography and for motivating the distinction between slowly moving amenity stocks and discrete entry shocks. We do not use those stock-based measures to define the headline treatment. In the main text they appear only as contextual measures or targeted confound checks; additional descriptive evidence is left to the appendix.

The main text now turns to event-time evidence around discrete entry shocks.


\section{Empirical Strategy and Main Results}

\subsection{Empirical Strategy}

Our main estimand is the reduced-form change in local log house prices following the first clean entry wave of food-and-drink establishments in an LSOA, relative to untreated LSOAs that remain at risk in the relevant event-time comparisons. It is best interpreted as a local capitalization estimand: it bundles direct amenity value, signaling, and related neighborhood revaluation, rather than isolating a single structural channel.

The central identification threat is endogenous siting. If businesses open where appreciation was already accelerating, post-entry coefficients will confound commercial entry with prior neighborhood change. To trace the dynamic price response while reducing that concern, we exploit the staggered timing of openings of food-and-drink establishments using EPC lodgement records for restaurants, cafes, drinking establishments, and takeaways. In the preferred specification, treatment begins in the first year an LSOA records at least two such EPC lodgements after a two-year lookback window with no prior eligible entry. Let $E_i$ denote the treatment year for LSOA $i$ (undefined for never-treated units). This clean-onset rule is intended to capture a discrete local entry wave rather than ongoing turnover. Our annual event-study specification is:

\begin{equation}
y_{it} = \alpha_i + \delta_t + \sum_{\substack{\tau = -5 \\ \tau \neq -1}}^{5} \beta_\tau \, \mathbf{1}\{t - E_i = \tau\} + \varepsilon_{it}
\label{eq:event_study}
\end{equation}

\noindent where $\mathbf{1}\{t - E_i = \tau\}$ equals one when LSOA $i$ is observed $\tau$ years relative to its treatment date, with $\tau = -1$ as the omitted reference period. The coefficients $\{\beta_\tau\}$ trace out the dynamic treatment effect, while the pre-period coefficients $\beta_{-5}, \ldots, \beta_{-2}$ provide the main visual and joint-test check on parallel trends.

\paragraph{Estimator comparisons.} Because staggered treatment timing can induce negative weighting in conventional TWFE \cite{goodman2021}, we treat the stacked design as the main estimator and compare it with a Sun and Abraham \cite{sun2021} interaction-weighted estimator. The point of this comparison is to check whether the post-entry path survives an alternative treatment-timing aggregation scheme, not to elevate a broad specification horse race into the main text.

\subsection{Main Event-Study Results}

Using EPC lodgement records, we construct an annual design for entry of food-and-drink establishments in which treatment begins when an LSOA first records at least two eligible lodgements after a two-year clean-onset window with no prior entry. The preferred stacked sample contains 657,197 LSOA-year observations, 774 treated-role LSOAs, and 4,459 control-role LSOAs across risk sets. Treatment is distributed across sixteen cohorts from 2008 to 2023, with the largest cohort in 2009 (178 treated units) and meaningful support throughout the subsequent decade. Appendix cohort tables report the full cohort composition.

\paragraph{Support across event time.} Table \ref{tab:support} summarizes the support behind the stacked sample. Treated observations are most numerous near event time and then decline as the window moves farther into the tails, but support remains substantial even at $\tau = 5$, where 546 treated observations from 11 contributing stacks remain in the estimating sample. This pattern matters for interpretation: medium-run coefficients are identified from fewer cohorts than the event-year effect, so later horizons should be read as economically informative but mechanically noisier.

\begin{table}[H]
\centering
\caption{Event-Time Support in the Stacked Sample}
\label{tab:support}
\begin{threeparttable}
\small
\begin{tabular}{@{}rccc@{}}
\toprule
\textbf{Event Time} & \textbf{Treated Obs.} & \textbf{Treated LSOAs} & \textbf{Contributing Stacks} \\
\midrule
$-5$ & 572 & 572 & 14 \\
$-4$ & 751 & 751 & 15 \\
$-3$ & 769 & 769 & 16 \\
$-2$ & 763 & 763 & 16 \\
$-1$ & 761 & 761 & 16 \\
$0$  & 756 & 756 & 16 \\
$1$  & 689 & 689 & 15 \\
$2$  & 648 & 648 & 14 \\
$3$  & 610 & 610 & 13 \\
$4$  & 586 & 586 & 12 \\
$5$  & 546 & 546 & 11 \\
\bottomrule
\end{tabular}
\begin{tablenotes}
\item Notes: Support counts from the preferred clean-onset stacked design for entry of food-and-drink establishments. ``Contributing stacks'' records how many cohort-specific risk sets identify each event-time coefficient.
\end{tablenotes}
\end{threeparttable}
\end{table}

\begin{figure}[H]
\centering
\includegraphics[width=0.92\textwidth]{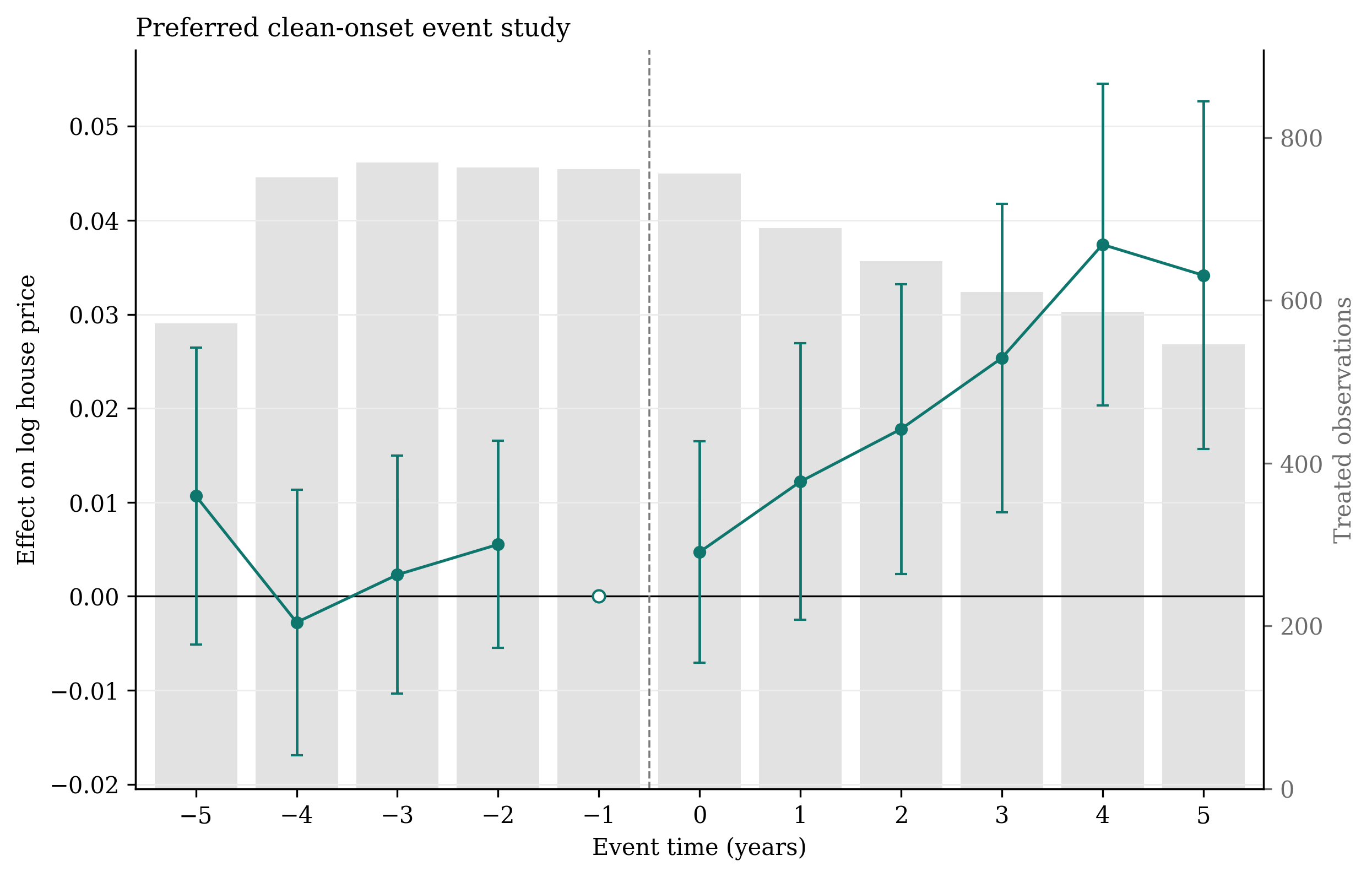}
\caption{Preferred clean-onset event-study estimates for entry of food-and-drink establishments. Point estimates are reported relative to the year before treatment ($\tau=-1$), which is shown as the open reference marker at zero, and plotted with 95\% confidence intervals. Gray bars report the number of treated observations contributing to each event-time coefficient. The estimated path rises gradually from near zero in the event year to roughly 3.4--3.7\% by years four and five after entry.}
\label{fig:event_study}
\end{figure}

\paragraph{Headline dynamic pattern.} Figure \ref{fig:event_study} reports the preferred clean-onset event-study path and overlays the event-time support behind each coefficient. Pre-treatment coefficients are small and imprecise, and the stacked joint pre-trend test does not reject ($p = 0.241$). The estimated effect is close to zero at event time and then turns positive: the stacked coefficient is 0.005 at $\tau = 0$, 0.012 at $\tau = 1$, 0.018 at $\tau = 2$, 0.037 at $\tau = 4$, and 0.034 by $\tau = 5$. Interpreted in logs, these magnitudes correspond to a gradual appreciation of roughly 0.5\% in the event year and about 3.4--3.7\% after four to five years. The pattern is notable for its slope as much as for its level: prices do not jump immediately, but instead drift upward as the new commercial entry becomes part of the local amenity bundle.

\paragraph{Estimator comparison and design diagnostics.} For the preferred clean-onset design, the Sun-Abraham interaction-weighted path closely tracks the stacked path in the post-treatment period. The Sun coefficients are 0.005 at $\tau = 0$, 0.012 at $\tau = 1$, 0.016 at $\tau = 2$, 0.036 at $\tau = 4$, and 0.033 at $\tau = 5$, and its joint pre-trend test is not rejected ($p = 0.472$). Table \ref{tab:main_design} places the preferred specification next to two nearby alternatives. Relative to the broader baseline threshold-2 design, clean-onset imposes more discipline on timing while preserving a similar medium-run magnitude. Relative to the neighbor-buffer variant, it avoids an extra layer of spatial pruning that mechanically strengthens post-treatment coefficients but yields tighter pre-trend $p$-values. We therefore retain the unbuffered clean-onset design as the headline specification.

\begin{table}[H]
\centering
\caption{Main Design Diagnostics for the Headline Event-Study Specification}
\label{tab:main_design}
\begin{threeparttable}
\footnotesize
\begin{tabular}{@{}lccccc@{}}
\toprule
\textbf{Specification} & \textbf{Treated-role LSOAs} & \textbf{Stacked pre-trend $p$} & \textbf{Sun pre-trend $p$} & \textbf{$\tau = 4$} & \textbf{$\tau = 5$} \\
\midrule
Baseline threshold-2 & 1,084 & 0.342 & 0.223 & 0.029 & 0.022 \\
Clean-onset threshold-2 & 774 & 0.241 & 0.472 & 0.037 & 0.034 \\
Clean-onset + neighbor buffer & 774 & 0.126 & 0.239 & 0.042 & 0.040 \\
\bottomrule
\end{tabular}
\begin{tablenotes}
\item Notes: Entries in the last two columns are log-point coefficients relative to $\tau = -1$. The baseline row uses the broader threshold-2 treatment without the clean-onset restriction. The preferred row uses the clean-onset threshold-2 treatment with a two-year lookback window and no prior eligible entry. The buffer row applies a first-order neighboring-control exclusion to the preferred design.
\end{tablenotes}
\end{threeparttable}
\end{table}

\subsection{Robustness Checks}

The main robustness question is whether the positive post-entry path survives plausible threats from spatial spillovers and concurrent redevelopment. Excluding first-order neighboring controls leaves the qualitative pattern intact and slightly strengthens post-treatment coefficients, but it also yields tighter pre-trend $p$-values. We therefore treat the buffer variant as a supporting robustness check rather than as a replacement for the headline design. Planning-start outcome diagnostics are reported in Appendix Table \ref{tab:appendix_checks} and are not used to select the preferred specification.


\section{Discussion}

\subsection{Interpretation}

The central empirical pattern is a gradual appreciation in local house prices after clean-onset entry of food-and-drink establishments. One interpretation is amenity capitalization: new venues increase neighborhood consumption value, signal local commercial upgrading, and alter expectations about future demand. The fact that the estimated response builds over several years is consistent with revaluation occurring as the new entry becomes embedded in the surrounding amenity bundle rather than through an immediate one-period jump.

\subsection{Limitations}

Several threats to identification merit discussion.

\textit{Endogenous siting.} Food-and-drink establishments may enter neighborhoods that were already on steeper appreciation paths. The clean-onset treatment improves pre-period behavior relative to broader entry definitions, but it does not by itself eliminate all concern about endogenous timing.

\textit{Treatment measurement.} EPC lodgements are an informative timing proxy, but they can reflect opening, refurbishment, reletting, or sale-related activity rather than a perfectly clean opening shock. Strengthening treatment validation is therefore an important next step for sharpening interpretation.

\textit{Spillovers and equilibrium responses.} Commercial entry may affect nearby untreated LSOAs, not just the treated unit itself. The local estimates reported here should therefore be interpreted as reduced-form neighborhood revaluation effects rather than as a full general-equilibrium accounting.

\textit{Displacement and external validity.} Our analysis captures observable commercial change and property-market dynamics, but not displacement directly \cite{freeman2005}. The evidence is also specific to Greater London during 2008--2023, a period shaped by post-crisis recovery, COVID-19, and uneven borough-level redevelopment.


\section{Conclusion}

This paper studies whether entry of food-and-drink establishments is followed by changes in neighborhood house prices in Greater London. In the preferred clean-onset event-study estimates, areas exposed to entry experience a gradual post-entry appreciation that reaches roughly 3.4--3.7\% after four to five years. The evidence is consistent with local amenity capitalization following visible commercial change, while also underscoring the need for continued falsification and treatment-validation work before treating the dynamic magnitudes as fully settled.


\appendix

\section*{Appendix}

\noindent This appendix retains only materials that directly support the preferred clean-onset event-study design.

\subsection*{Sun--Abraham Cross-Check}

\begin{figure}[H]
\centering
\includegraphics[width=0.92\textwidth]{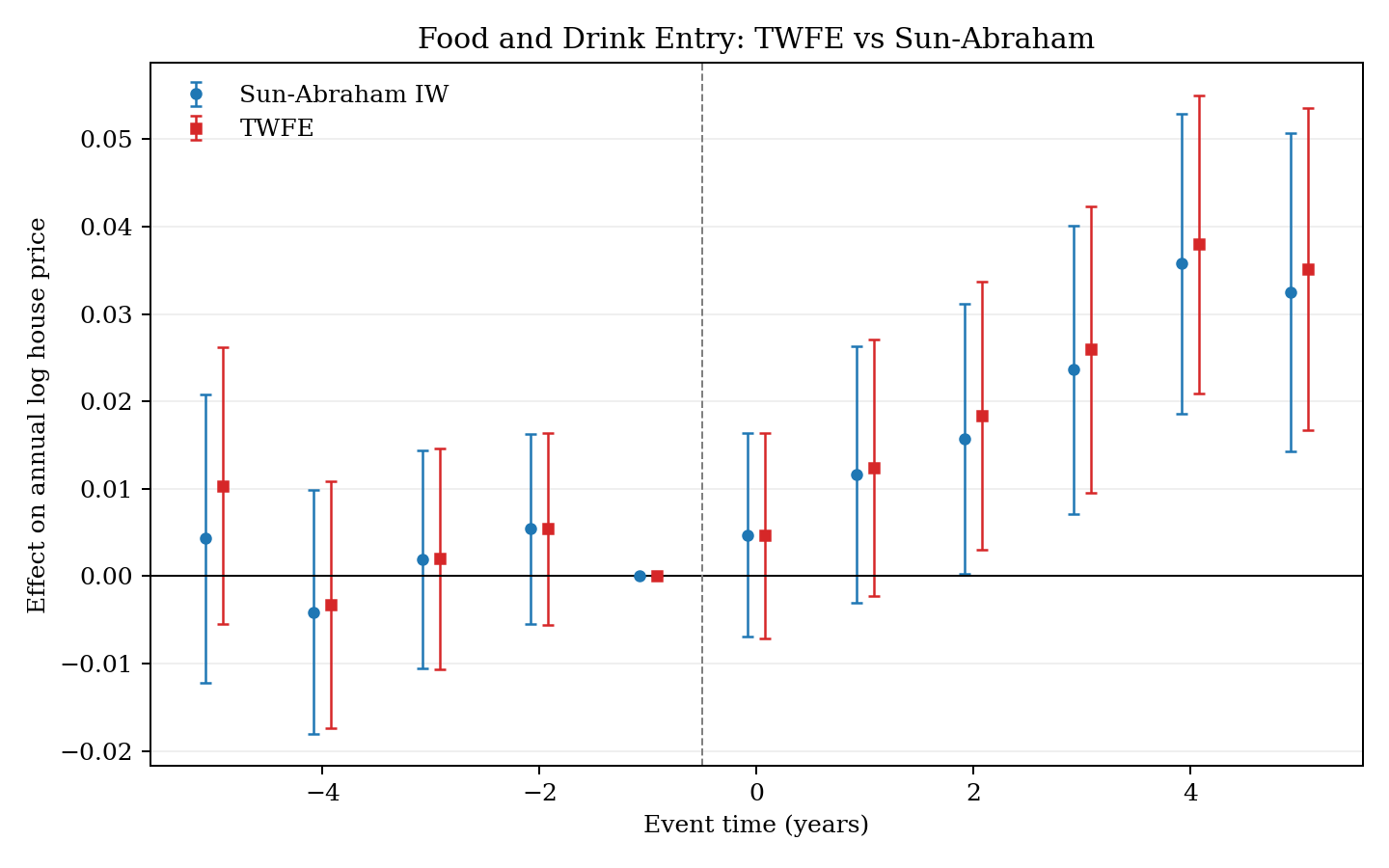}
\caption{Sun--Abraham estimates for the preferred clean-onset specification for entry of food-and-drink establishments. The interaction-weighted path closely tracks the stacked estimates in the post-treatment period and does not reject a joint pre-trend test.}
\label{fig:sun_compare}
\end{figure}

\subsection*{Cohort Composition}

\begin{table}[H]
\centering
\caption{Cohort Composition of the Preferred Clean-Onset Design for Entry of Food-and-Drink Establishments}
\label{tab:cohort_summary}
\begin{threeparttable}
\small
\begin{tabular}{@{}rrrr@{}}
\toprule
\textbf{Cohort Year} & \textbf{Treated Units} & \textbf{Control Units} & \textbf{Stack Rows} \\
\midrule
2008 & 16 & 4,459 & 39,597 \\
2009 & 178 & 4,437 & 45,431 \\
2010 & 69 & 4,411 & 48,513 \\
2011 & 38 & 4,363 & 47,635 \\
2012 & 42 & 4,320 & 47,135 \\
2013 & 29 & 4,271 & 46,325 \\
2014 & 22 & 4,237 & 45,721 \\
2015 & 26 & 4,213 & 45,457 \\
2016 & 48 & 4,169 & 45,225 \\
2017 & 43 & 4,126 & 44,676 \\
2018 & 49 & 4,057 & 43,761 \\
2019 & 34 & 4,057 & 39,568 \\
2020 & 24 & 4,057 & 35,436 \\
2021 & 44 & 4,057 & 31,562 \\
2022 & 43 & 4,057 & 27,504 \\
2023 & 69 & 4,057 & 23,651 \\
\bottomrule
\end{tabular}
\begin{tablenotes}
\item Notes: Each row reports one cohort-specific risk set in the preferred clean-onset stacked design for entry of food-and-drink establishments. ``Stack rows'' counts all treated and control observations contributed by that cohort-specific stack.
\end{tablenotes}
\end{threeparttable}
\end{table}

\subsection*{Additional Event-Study Checks}

\begin{table}[H]
\centering
\caption{Supporting Event-Study Checks for the Preferred Design}
\label{tab:appendix_checks}
\begin{threeparttable}
\small
\begin{tabular}{@{}lccc@{}}
\toprule
\textbf{Check} & \textbf{Outcome} & \textbf{Pre-trend $p$} & \textbf{$\tau = 4 / \tau = 5$} \\
\midrule
Neighbor-buffer exclusion & Log house price & 0.126 & 0.042 / 0.040 \\
Planning confound check & asinh residential planning-start GIA & 0.265 & 0.094 / 0.041 \\
\bottomrule
\end{tabular}
\begin{tablenotes}
\item Notes: Both rows use the preferred clean-onset treatment definition for entry of food-and-drink establishments. The buffer row excludes first-order neighboring controls. The planning row re-estimates the event-study on residential planning-start GIA gained to assess whether the main price response is mechanically tracking redevelopment starts.
\end{tablenotes}
\end{threeparttable}
\end{table}



\begin{thebibliography}{99}

\bibitem{zukin2009}
Zukin, S., Trujillo, V., Frase, P., Jackson, D., Recuber, T., \& Walker, A. (2009). New retail capital and neighborhood change. \emph{City \& Community}, 8(1), 47--64.

\bibitem{smith1979}
Smith, N. (1979). Toward a theory of gentrification: A back to the city movement by capital, not people. \emph{Journal of the American Planning Association}, 45(4), 538--548.

\bibitem{ley2003}
Ley, D. (2003). Artists, aestheticisation and the field of gentrification. \emph{Urban Studies}, 40(12), 2527--2544.

\bibitem{zukin1982}
Zukin, S. (1982). \emph{Loft Living: Culture and Capital in Urban Change}. Johns Hopkins University Press.

\bibitem{butler2003}
Butler, T. (2003). Living in the bubble: Gentrification and its `others' in North London. \emph{Urban Studies}, 40(12), 2469--2486.

\bibitem{guerrieri2013}
Guerrieri, V., Hartley, D., \& Hurst, E. (2013). Endogenous gentrification and housing price dynamics. \emph{Journal of Public Economics}, 100, 45--60.

\bibitem{dube2014}
Dub\'{e}, J., Legros, D., Th\'{e}riault, M., \& Des Rosiers, F. (2014). A spatial difference-in-differences estimator to evaluate the effect of change in public mass transit systems on house prices. \emph{Transportation Research Part B}, 64, 24--40.

\bibitem{heckert2012}
Heckert, M., \& Mennis, J. (2012). The economic impact of greening urban vacant land: A spatial difference-in-differences analysis. \emph{Environment and Planning B}, 39(6), 1105--1121.

\bibitem{wang2022}
Wang, J., \& Deng, K. (2022). Impact and mechanism analysis of smart city policy on urban innovation: Evidence from China. \emph{Economic Analysis and Policy}, 73, 574--592.

\bibitem{dong2021}
Dong, L., Du, R., Kahn, M., Ratti, C., \& Zheng, S. (2021). ``Ghost cities'' versus boom towns: Do China's high-speed rail new towns thrive? \emph{Regional Science and Urban Economics}, 89, 103682.

\bibitem{reades2019}
Reades, J., De Souza, J., \& Hubbard, P. (2019). Understanding urban gentrification through machine learning. \emph{Urban Studies}, 56(5), 922--942.

\bibitem{atkinson2000}
Atkinson, R. (2000). Measuring gentrification and displacement in Greater London. \emph{Urban Studies}, 37(1), 149--165.

\bibitem{liu2023}
Liu, C., \& Bardaka, E. (2023). Transit-induced commercial gentrification: Causal inference through a difference-in-differences analysis of business microdata. \emph{Transportation Research Part A}, 175, 103788.

\bibitem{hackworth2002}
Hackworth, J., \& Smith, N. (2002). The changing state of gentrification. \emph{Tijdschrift voor Economische en Sociale Geografie}, 92(4), 464--477.

\bibitem{glass1964}
Glass, R. (1964). Introduction to London: Aspects of change. \emph{Centre for Urban Studies}.

\bibitem{barton2016}
Barton, M. (2016). An exploration of the importance of the strategy used to identify gentrification. \emph{Urban Studies}, 53(1), 92--111.

\bibitem{hillsdon2006}
Hillsdon, M., Panter, J., Foster, C., \& Jones, A. (2006). The relationship between access and quality of urban green space with population physical activity. \emph{Public Health}, 120(12), 1127--1132.

\bibitem{wolch2014}
Wolch, J. R., Byrne, J., \& Newell, J. P. (2014). Urban green space, public health, and environmental justice. \emph{Landscape and Urban Planning}, 125, 234--244.

\bibitem{greco2019}
Greco, S., Ishizaka, A., Tasiou, M., \& Torrisi, G. (2019). On the methodological framework of composite indices. \emph{Social Indicators Research}, 141(1), 61--94.

\bibitem{decancq2013}
Decancq, K., \& Lugo, M. A. (2013). Weights in multidimensional indices of wellbeing: An overview. \emph{Econometric Reviews}, 32(1), 7--34.

\bibitem{anselin2006}
Anselin, L. (2006). Spatial econometrics. In \emph{Palgrave Handbook of Econometrics}, Vol. 1, pp. 901--969.

\bibitem{goodman2021}
Goodman-Bacon, A. (2021). Difference-in-differences with variation in treatment timing. \emph{Journal of Econometrics}, 225(2), 254--277.

\bibitem{sun2021}
Sun, L., \& Abraham, S. (2021). Estimating dynamic treatment effects in event studies with heterogeneous treatment effects. \emph{Journal of Econometrics}, 225(2), 175--199.

\bibitem{freeman2005}
Freeman, L. (2005). Displacement or succession? Residential mobility in gentrifying neighborhoods. \emph{Urban Affairs Review}, 40(4), 463--491.

\end{thebibliography}
\end{document}